\documentclass{emulateapj}
\def\deg{\ifmmode^\circ\else$^\circ$\fi}

\def\gs{{_>\atop^{\sim}}}

\def\lsun{L$_{\odot}$}

\def\arcs{\ifmmode {''}\else $''$\fi}
\def\arcm{\ifmmode {'}\else $'$\fi}
\def\parcs{\sa=.07em \sb=.03em
     \ifmmode $\rlap{.}$^{\scriptscriptstyle\prime\kern -\sb\prime}$\kern -\sa$
     \else \rlap{.}$^{\scriptscriptstyle\prime\kern -\sb\prime}$\kern -\sa\fi}
\def\parcm{\sa=.08em \sb=.03em
     \ifmmode $\rlap{.}\kern\sa$^{\scriptscriptstyle\prime}$\kern-\sb$
     \else \rlap{.}\kern\sa$^{\scriptscriptstyle\prime}$\kern-\sb\fi}

\def\Msun{M$_{\odot}$}

\def\spose#1{\hbox to 0pt{#1\hss}}
\def\simlt{\mathrel{\spose{\lower 3pt\hbox{$\mathchar"218$}}
     \raise 2.0pt\hbox{$\mathchar"13C$}}}
\def\simgt{\mathrel{\spose{\lower 3pt\hbox{$\mathchar"218$}}
     \raise 2.0pt\hbox{$\mathchar"13E$}}}
\def\lsim{\rlap{$<$}{\lower 1.0ex\hbox{$\sim$}}}
\def\gsim{\rlap{$>$}{\lower 1.0ex\hbox{$\sim$}}}

\begin{document}
\slugcomment{ApJSup accepted (Spitzer Special Issue)}
\shorttitle{Spitzer/IRS Observations of ULIRGs}
\shortauthors{Armus et al.}

\title{Observations of Ultraluminous Infrared Galaxies with the Infrared Spectrograph 
on the Spitzer Space Telescope: Early Results on Mrk 1014, Mrk 463, and UGC 5101\altaffilmark{1}}

\author{L. Armus\altaffilmark{2},
V. Charmandaris\altaffilmark{3,6}, H.W.W. Spoon\altaffilmark{3},
J.R. Houck\altaffilmark{3}, B.T. Soifer\altaffilmark{2},
B.R. Brandl\altaffilmark{4}, P.N. Appleton\altaffilmark{2}, H.I. Teplitz\altaffilmark{2},
S.J.U. Higdon\altaffilmark{3}, D.W. Weedman\altaffilmark{3}, D. Devost\altaffilmark{3},
P.W. Morris\altaffilmark{2},
K.I. Uchida\altaffilmark{3}, J. van Cleve\altaffilmark{5}, D.J. Barry\altaffilmark{3},
G.C. Sloan\altaffilmark{3},
C.J. Grillmair\altaffilmark{2},
M.J. Burgdorf\altaffilmark{2},
S.B. Fajardo-Acosta\altaffilmark{2}, J.G. Ingalls\altaffilmark{2}, J. Higdon\altaffilmark{3}
L. Hao\altaffilmark{3}, J. Bernard-Salas\altaffilmark{3}, T. Herter\altaffilmark{3},
J. Troeltzsch\altaffilmark{5}, B. Unruh\altaffilmark{5},
M. Winghart\altaffilmark{5}}

\altaffiltext{1}{based on observations obtained with the Spitzer Space Telescope, which is
operated by the Jet Propulsion Laboratory, California Institute of Technology, under NASA 
contract 1407}
\altaffiltext{2}{Spitzer Science Center, MS 220-6, Caltech, Pasadena, CA 91125}
\altaffiltext{3}{Astronomy Department, Cornell University, Ithaca, NY 14853}
\altaffiltext{4}{Leiden University, 2300 RA Leiden, The Netherlands}
\altaffiltext{5}{Ball Aerospace \& Technologies Corp., 1600 Commerce St., Boulder CO, 80301}
\altaffiltext{6}{Chercheur Associ\'e, Observatoire de Paris, F-75014, Paris, France}
\altaffiltext{7}{The IRS was a collaborative venture between Cornell University and 
Ball Aerospace Corporation funded by NASA through the Jet Propulsion Laboratory and
the Ames Research Center}

\begin{abstract}

We present spectra taken with the Infrared
Spectrograph$^{7}$ on Spitzer covering the $5-38\mu$m
region of three Ultraluminous Infrared Galaxies (ULIRGs):
Mrk 1014 ($z=0.163$), and Mrk 463 ($z=0.051$), and UGC 5101
($z=0.039$).
The continua of UGC 5101 and Mrk 463
show strong silicate absorption suggesting significant optical
depths to the nuclei at $10\mu$m.  UGC 5101 also shows the clear
presence of water ice in absorption.  PAH emission features are
seen in both Mrk 1014 and UGC 5101, including the $16.4\mu$m
line in UGC 5101.  The fine structure lines are consistent
with dominant AGN power sources in both Mrk 1014 and Mrk 463.
In UGC 5101 we detect the [NeV] $14.3\mu$m emission line
providing the first direct evidence for a buried AGN in the
mid-infrared.  The detection of the $9.66\mu$m and $17.03\mu$m
H$_{2}$ emission lines in both UGC 5101 and Mrk 463 suggest
that the warm molecular gas accounts for 22\%
and 48\% of the total molecular gas masses in these galaxies.

\end{abstract}


\section{Introduction}

Ultraluminous Infrared Galaxies (ULIRGs), i.e. those galaxies
with infrared luminosity L$_{IR} \gs 10^{12}$\lsun, have
the power output of quasars yet emit nearly all of their
energy in the mid and far-infrared part of the spectrum.
Most ULIRGs are found in interacting and merging systems
(e.g. Armus, Heckman \& Miley 1987; Sanders, et al. 1988a;
Murphy, et al. 1996), where the merger has driven
gas and dust towards the remnant nucleus, fueling a massive
starburst, and either creating or fueling a nascent AGN (Mihos
\& Hernquist 1996).  ULIRGs are rare in the local Universe,
comprising only $3\%$ of the IRAS Bright Galaxy Survey (Soifer
et al. 1987), yet at $z > 2-3$, ULIRGs may account for the bulk
of all star-formation activity and dominate the far-infrared background
(e.g. Blain, et al. 2002).

Observations with the ISO satellite greatly expanded our
understanding of the mid-infrared spectra of ULIRGs (e.g.,
Genzel et al. 1998; Lutz et al. 1999; Rigopoulou et al. 1999;
Sturm et al. 2002; Tran et al. 2001).  Diagnostic diagrams
based on fine structure and aromatic emission features (UIB's,
or PAH's) allowed some ULIRGs to be classified according to
their dominant ionization mechanism.  However, the complexities
of the ULIRG spectra, the fact that many are likely composite
AGN and starburst sources, and the limitations in sensitivity
of the ISO spectrometers, left many ULIRGs, even at relatively
low redshift, beyond the reach of these methods until now.

In order to adequately sample the local ULIRG population,
we are obtaining mid-infrared spectra of a large number ($ > 100$) of
ULIRGs having $0.02 < z < 0.93$ with the Infrared 
Spectrograph (IRS) 
on Spitzer, as part of the IRS guaranteed
time program.  These sources are chosen from the complete
Bright Galaxy Sample (Soifer et al. 1987), the 1-Jy (Kim \&
Sanders 1998) and 2-Jy (Strauss et al. 1992) samples, and
the FIRST/IRAS radio-far-IR sample of Stanford et al. (2000).
In this letter, we present the first results from this program,
focusing on three nearby ULIRGs (UGC 5101, Mrk 463, and Mrk
1014) whose spectra reflect the range of properties we expect
from the sample as a whole.

UGC 5101 ($z=0.039$) has a single, very red nucleus within
a disturbed morphology suggestive of a recent interaction
and merger.  Optically, UGC 5101 is classified as a LINER
(Veilleux et al. 1995).  It has a high brightness temperature
(T$>10^{7}$K) radio nucleus at 1.6 GHz which is resolved
with the VLBA (Lonsdale et al. 1995).  ISO SWS and PHT-S
spectroscopy (Genzel el al. 1998) indicate a powerful,
circumnuclear starburst.  Ground-based, high-resolution mid-IR
imaging (Soifer et al. 2000) indicates that $\approx 60$\% of
the IRAS flux at $12\mu$m comes from the central four arcsec,
and that nearly half of this arises in an unresolved core.
Based upon its IRAS colors, UGC 5101 is classified as a cold,
starburst-dominated, far-infrared source.  However, XMM
data indicate an obscured, but luminous, hard x-ray source
with L$_{x}$(2-10 keV)$\sim 5\times 10^{42}$ erg s$^{-1}$
and L$_{x}$(2-10 keV)/L$_{IR} \sim 0.002$ suggestive of a
buried AGN (Imanishi, et al. 2003).  Here, we present the
first direct infrared evidence for an AGN in UGC 5101.

Mrk463 ($z=0.0508$) is a merging system with two nuclei
separated by about four arcseconds (Mazzarella, et al. 1991).
Both nuclei have Seyfert 2 optical spectra (Shuder \& Osterbrock
1981), but broad lines are seen in scattered optical (Miller
\& Goodrich 1990) and direct near-infrared light (Goodrich,
Veilleux \& Hill 1994; Veilleux et al. 1997).  The eastern
nucleus (Mrk463e) is much redder ($V-K = 6.8$ mag), and has
a luminous, steep-spectrum radio core (Mazzarella et al.).
Although the far-infrared luminosity of Mrk 463 ($5\times
10^{11}$\lsun) is slightly less than the canonical ULIRG cutoff
of $10^{12}$\lsun, the bolometric luminosity of this system
is very high, and we refer to it as a ULIRG for the remainder
of this paper.

Mrk 1014 ($z=0.1631$) is a radio-quiet, infrared luminous
QSO with broad optical emission lines (FWHM H$\beta >
4000$km s$^{-1}$) and twin tidal tails indicative of a
recent interaction and merger (MacKenty \& Stockton 1984).
Both Mrk 463 and Mrk 1014 are warm, far-infared sources with
S$_{25}$/S$_{60} = 0.74$ and 0.27, respectively (Sanders,
et al. 1988b).

Throughout the paper, we will adopt a flat, $\Lambda$-dominated
Universe ($H_0 = 70$ km s$^{-1}$\ Mpc$^{-1}$, $\Omega_M=0.3$,
$\Omega_{\Lambda}=0.7$).  The luminosity distances to UGC 5101, Mrk
463, and Mrk 1014 are then 170 Mpc, 223 Mpc, and 774 Mpc,
and one arcsec subtends 0.76, 0.98, and 2.77 kpc in projection,
respectively.

\section{Observations}

Details of the observations are given in Table 1.  The IRS is fully
described in Houck et al. (2004).  All three ULIRGs were observed in
the two low-resolution ($64 < R < 128$; Short-Low and Long-Low or SL
\& LL) and two high-resolution ($R\sim600$; Short-High and Long-High
or SH \& LH) IRS modules, using the Staring Mode Astronomical
Observing Template (AOT).  High accuracy blue peak-ups were performed
on nearby 2MASS stars before offsetting to the target galaxies.  For
Mrk463, the eastern nucleus, Mrk463e, was centered in the IRS slits in
all cases.  While the separation of the nuclei is comparable to or
less than the widths of the IRS slits, we expect Mrk463e to dominate
at all IRS wavelengths.

\begin{deluxetable}{cccc}
\tabletypesize{\scriptsize}
\tablecaption{Observation Log\label{tbl-1}}
\tablehead{
\colhead {} &
\colhead{\bf{UGC 5101}} &
\colhead{\bf{Mrk 1014}} &
\colhead{\bf{Mrk 463e}} 
}

\startdata
date&15 Nov 2003&7 Jan 2004&7 Jan 2004\\
 &(23 Mar 2004)& &\\
PU target&BD$+62$ 1078&HD121829&HD12382\\
SL1&6x14sec&6x14sec&6x14sec\\
SL2&6x14sec&6x14sec&6x14sec\\
LL1&4x30sec&4x30sec&6x14sec\\
LL2&4x30sec&4x30sec&6x14sec\\
SH&2x30sec&12x30sec&12x30sec\\
LH&2x60sec&8x60sec&8x60sec\\
\enddata

\tablecomments{IRS observation details, including dates, peakup
  target, and integration times, are given for each galaxy.  For each
  slit we list the cycles and ramp durations, where 6x14sec indicates
  six cycles of 14sec ramps, including both nod positions.  The SL and
  LL observations for UGC 5101 were performed on 23 March 2004, while
  the SH and LH observations were performed on 15 November 2003.}
                                                                                                           
\end{deluxetable}

\section{Data Reduction and Analysis}

All spectra were reduced using the IRS pipeline at the Spitzer Science
Center.  This reduction includes ramp fitting, dark sky subtraction,
droop correction, linearity correction, flat fielding, and wavelength
and flux calibration (see Chapter 7 of the Spitzer Observers Manual
and Decin et al. 2004 for details).  The SL and LL data have had local
background light subtracted, by differencing the two nod positions
along the slit, before spectral extraction.  As a final step, we have
normalized the SL and LL 1D spectra upwards to match the 12 and
$25\mu$m IRAS FSC data (Moshir et al. 1990).  The final, average SL
and LL spectra are displayed in Fig.1.

 \begin{figure}
 \epsscale{1.1}
 \plotone{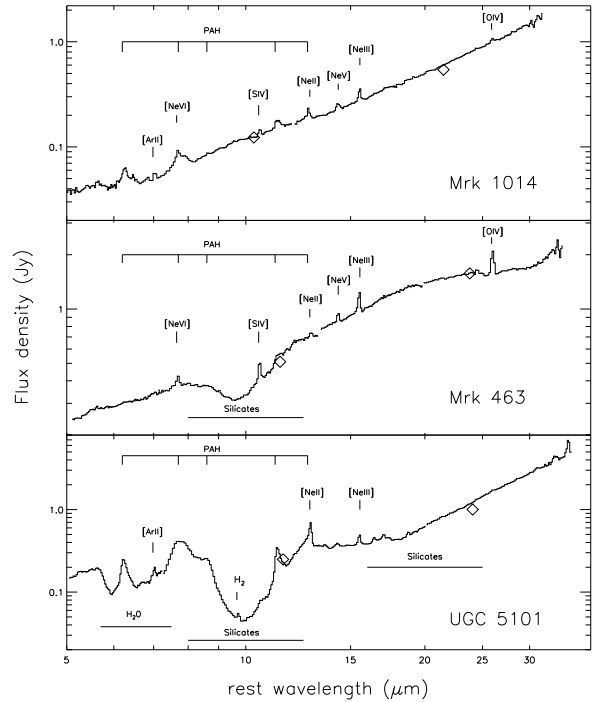}
 \caption{IRS Short-Low \& Long-Low spectra of Mrk 1014, Mrk 463e and UGC 5101.  Prominent 
 emission lines and absorption bands (the latter indicated by horizontal bars) are marked.
 Open diamonds are the
 IRAS $12\mu$m and $25\mu$m points.}  
 \end{figure}

 Since the SH and LH slits are too small for on-slit background
 subtraction, we have subtracted the expected background flux through
 each slit based on the model of Reach, et al. (2004).  The SH and LH
 spectra were then scaled, on an order by order basis, to the
 corresponding low-resolution data.  The average SH and LH spectra are
 shown in Figs 2 \& 3, respectively.  While most high-res orders line
 up very well, there are slight offsets and residual curvature still
 visible in some orders (e.g. the bluest SH orders of Mrk 1014 and at
 rest wavelengths of $12\mu$m and $22\mu$m).  Noisy areas at the red
 end of the SH and LH orders in order overlap regions are not shown or
 used in the fitting process. In SH, these noisy areas amount to
 typically 10-30 pixels at the red end of orders 13-20, corresponding
 to an area of decreased responsivity on the array.

 \begin{figure}
 \epsscale{1.1}
 \plotone{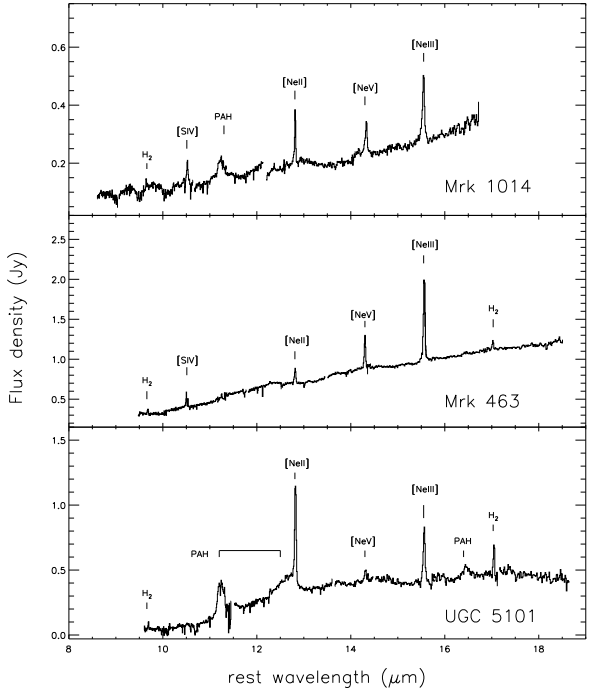}
 \caption{IRS Short-High spectra of Mrk 1014, Mrk 463e and UGC 5101.  Prominent
 emission lines are marked.}
 \end{figure}

 \begin{figure}
 \epsscale{1.1}
 \plotone{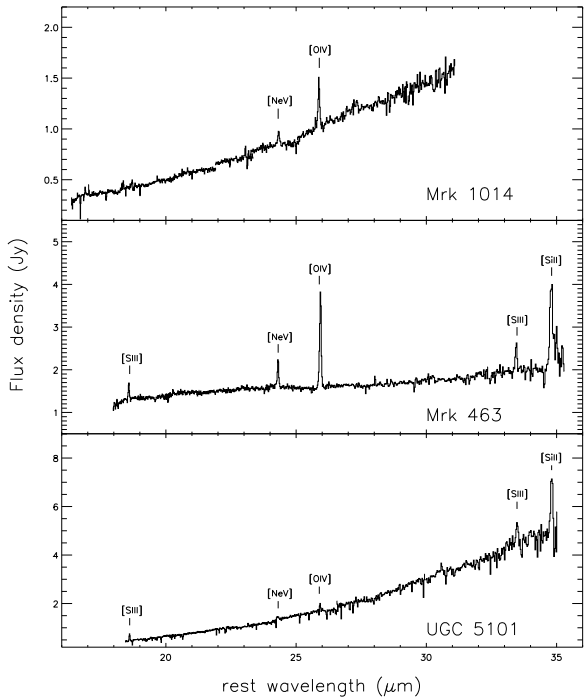}
 \caption{IRS Long-High spectra of Mrk 1014, Mrk 463e and UGC 5101.  Prominent
 emission lines are marked.}
 \end{figure}

\section{Results}

\subsection{Continuum \& Absorption Features}

The continua of Mrk 1014, Mrk 463e and UGC 5101 are strikingly
different. The low-resolution spectrum of UGC 5101 is a blend
of a water ice (most easily seen as a strong absorption under
the $6.2\mu$m PAH emission feature), and silicate-absorbed
(at $9.7\mu$m and $18\mu$m) starburst spectrum (see Fig. 1).
Much of the silicate absorption is filled in by the flanking
PAH emission features.  Adopting a smooth continuum anchored
at $5.3-5.6\mu$, $14\mu$m and $34\mu$m, we derive optical
depths of $\tau_{6} =0.75$, $\tau_{9.7} \ge2$, and $\tau_{18} \ge0.4$.
The extinction to the central source is thus at least
A$_{V} = 15-35$ mag.  As suggested by Spoon et al. (2002),
the water ice features may indicate the presence of shielded
molecular clouds along the line of sight to the nucleus.
While the $5.7-7.5\mu$m
absorption is dominated by water ice, the structure visible from
$6.8-7.5\mu$m is consistent with hydrocarbon absorption at the
level ($\tau_{6.85} \sim 0.16$) expected from the $3.4\mu$m
band (Imanishi et al. 2001), assuming an optical depth ratio
of $\tau_{6.85}$/$\tau_{3.4} = 0.238$ (as measured for SgrA*
by Chiar et al. 2000).

Mrk 1014 shows no obvious silicate absorption, suggesting a
rather clean line of sight at $10\mu$m to the Seyfert 1 nucleus.
Mrk 463e is intermediate in its continuum properties between UGC
5101 and Mrk 1014.  The most obvious continuum feature is the
silicate absorption at $9.7\mu$m. Adopting a power-law continuum
from $7-14\mu$m, we derive an optical depth $\tau_{9.7} =
0.46$, corresponding to an A$_{V} =3-8$ mag.

\subsection{Emission Features}

At high resolution, the IRS spectra of Mrk 1014, Mrk 463e and
UGC 5101 are dominated by unresolved atomic, fine-structure
lines of Ne, O, Si, and S, covering a large range in ionization
potential (see Figs. 2 \& 3 and Table 2).  Ratios of these
lines can be used to gauge the dominant ionizing source -
whether it be hot stars or an active nucleus.  Some features,
e.g. the [NeV] lines at 14.3 and $24.3\mu$m, imply the presence
of an AGN by their very detection in an integrated spectrum of
a galaxy, since it takes 97.1 eV to ionize [NeIV], and this
is too large to be produced by OB stars.  The same is not
true for [OIV], since it takes only 55 eV to ionize [OIII],
and in fact the $25.9\mu$m line has been seen in a number of
starburst galaxies (e.g., Lutz et al. 1998).

The [NeV]14.3/[NeII]12.8 and [OIV]25.9/[NeII]12.8 line flux
ratios in Mrk 1014 and Mrk 463e suggest that nearly all the
ionizing flux in these sources ($ > 80-90$\% based upon the
simple scaling models in Sturm et al. 2002) comes from the
active nuclei.  In addition, the [NeII]12.8/[NeIII]15.6,
[NeV]14.3/[NeIII]15.6, and [NeV]14.3/[NeV]24.3 line flux
ratios suggest that the electron densities are low ($10^{2}
- 10^{3}$ cm$^{-3}$) and that the ionization parameters and
spectral indices are in the ranges of $-1 <$log U$< -2$ and
$\alpha \le -1$ for both sources (Voit 1992, Sturm et al. 2002).
Weak [NeVI] $7.65\mu$m emission is also present in Mrk 1014 and
Mrk 463e (see Fig.1), but it is blended with the $7.7\mu$m PAH
feature and possibly even Pf-$\alpha$.  A proper separation
of these components can only be done with IRS high-res data
for galaxies with $z \ge 0.33$.

\begin{deluxetable}{cccc}
\tabletypesize{\scriptsize}
\tablecaption{Emission Features\label{tbl-1}}
\tablehead{
\colhead{} &
\colhead{\bf{UGC 5101}} &
\colhead{\bf{Mrk 1014}} &
\colhead{\bf{Mrk 463e}} \\
\colhead{\bf{Line $\mu$m}} &
\colhead{} &
\colhead{}&
\colhead{} 
}

\startdata
PAH 6.2&190(11)&14.4(1.8)\\
 &0.258&0.048\\
$[ArII]$ 6.98&15.2(2.5)$^{a}$\\
 &0.018\\
$[NeVI]$ 7.65& &5.5(0.5)$^{b}$&18.7(2.1)$^{b}$\\
 & &0.022&0.009\\
PAH 7.7&560(26)&32(5)$^{b}$&33(6)$^{b}$\\
 &0.575&0.135&0.029\\
H$_{2}$ 9.66&3.7(0.7)&1.6(0.5)&4.8(0.8)\\
 &0.034&0.007&0.003\\
$[SIV]$ 10.51&$\le 2.4$&7.9(0.9)&9.0(1.2)\\
 & &0.033&0.010\\
PAH 11.3&106(10)&4.6(0.8)\\
 &0.284&0.017\\
$[NeII]$ 12.81&55.2(2.5)&7.8(0.8)&11.6(0.7)\\
 &0.080&0.031&0.010\\
$[NeV]$ 14.3&5.2(0.7)&6.8(0.6)&18.3(0.8)\\
 &0.009&0.029&0.015\\
$[NeIII]$ 15.55&23.9(1.4)&13.3(1.1)&51.8(2.5)\\
 &0.050&0.054&0.052\\
PAH 16.4&14.5(1.9)\\
 &0.031\\
H$_{2}$ 17.03&7.2(1.4)& &3.8(0.5)\\
 &0.015& &0.004\\
$[SIII]$ 18.71&9.8(0.7)& &15.0(1.6)\\
 &0.025& &0.013\\
$[NeV]$ 24.31&4.9(1.0)&5.1(0.5)&20.4(1.7)\\
 &0.009&0.017&0.030\\
$[OIV]$ 25.89&5.5(1.4)&13.5(1.5)&72.3(1.2)\\
 &0.010&0.040&0.117\\
$[SIII]$ 33.48&13.0(2.5)& &13.5(1.6)\\
 &0.016& &0.033\\
$[SiII]$ 34.81&36(6)& &30.3(5.9)\\
 &0.047& &0.076\\
\enddata

\tablecomments{For each line
we give the central wavelength in microns, the flux, in
units of $10^{-21}$Wcm$^{-2}$ and the equivalent width, in
microns, directly below. Formal uncertainties in the line
fits (all single Gaussians except where noted) are listed in
parentheses next to the fluxes.  However, true uncertainties
in the absolute line fluxes are generally $\sim20-25$\%.
All lines were measured using the SMART spectral reduction
package (Higdon et al. 2004).  $^{a}$The [ArII]6.98 line is
blended with the H$_{2}$ S(5) line in UGC 5101.  $^{b}$The
[NeVI]7.65 line is blended with the $7.7\mu$m PAH feature in
Mrk 1014 and Mrk 463e, and a two-component Gaussian has been
used to separate the unresolved and broad components.}

\end{deluxetable}

In UGC 5101 we have clearly detected a [NeV] $14.3\mu$m line
with a flux of $5.2(\pm 0.7)\times 10^{-21}$W cm$^{-2}$
indicating a buried AGN.  The measured [NeV]14.3
flux is approximately a factor of three below the upper
limit set by Genzel et al. (1998).  We also detect the
[OIV] $25.89\mu$m line with a flux of $5.5(\pm 1.4)\times
10^{-21}$W cm$^{-2}$.  A faint [NeV] $24.3\mu$m line is also
visible in the LH spectrum.  The [SIII]18.7/[SIII]33.4 and the
[NeV]14.3/[NeV]24.3 line flux ratios imply electron densities
at or below $10^{2}$cm$^{-3}$, and the [NeIII]15.5/[NeII]12.8
line flux ratio implies a moderate excitation starburst
(Verma et al. 2003).  While the [NeV]14.3/[NeII]12.8 and the
[OIV]25.9/[NeII]12.8 line flux ratios ($\sim 0.1$ in each
case) are consistent with an AGN contribution of $< 10$\% to
the total luminosity in this source (Sturm et al. 2002), the
large optical depth to the nucleus, as evidenced by the deep
silicate absorption, leaves open the possibility that the true
contribution of the AGN to the bolometric power output in UGC
5101 may be larger than revealed by the mid-IR emission lines.

The  H$_{2}$ S(3) $9.66\mu$m and S(1) $17.04\mu$m pure
rotational lines from warm molecular gas are seen in both UGC
5101 and Mrk 463e.  In UGC 5101 the S(3)/S(1) line flux ratio
is about 0.5, while in Mrk 463e it is $\sim 1.3$, implying
warm gas temperatures (assuming LTE) of approximately 300K
and 400K, respectively.  If the S(1) line emission is from an
unresolved source in both cases, there is $\sim1.1$ and 0.5
$\times 10^{9}$\Msun of warm molecular gas in UGC 5101 and
Mrk 463e, respectively. These warm molecular gas estimates are
22\% and 48\%, respectively, of the total H$_{2}$ masses (warm
plus cold) in these galaxies for $\alpha = 0.8$ M$_{\odot}$
(K km s$^{-1}$pc$^{2}$)$^{-1}$ (Solomon et al 1997; Evans et
al. 2002).

We detect the 6.2, 7.7, 11.3, and $12.7\mu$m PAH emission
features in UGC 5101, and all but the $12.7\mu$m feature in
Mrk 1014 (see Figs. 1 \& 2).  A weak $7.7\mu$m feature may
be present in the SL spectrum of Mrk 463e.  In addition, we
detect the $16.4\mu$m PAH feature (e.g., Moutou et al. 2000)
in UGC 5101.  To our knowledge, this is the first detection of
this feature in a ULIRG, although it is seen in some 
nearby starburst galaxies (Sturm et al. 2000; Smith et al. 2004).
The $7.7\mu$m line-to-continuum (l/c) ratios we measure for Mrk 1014, Mrk 463e
and UGC 5101 are 0.25, 0.05, and 1.1, respectively.  An l/c $\ge 1.0$
is usually taken to imply a dominant starburst contribution to the
mid-infrared flux (Rigopoulou, et al. 1999).  The relatively
strong $7.7\mu$m, and $11.3\mu$m features in Mrk 1014, may
indicate that a circumnuclear starburst is present in this
Seyfert 1 galaxy.



\acknowledgements

We would like to thank Aaron Evans, Bruce Draine, Jacqueline
Keane, Bill Reach, J.D. Smith, Jason Surace and Mark Voit
for many helpful discussions.  We would also like to thank
the referee, Eckhard Sturm, for a careful reading of the
manuscript.  It is also a pleasure to thank the countless people
at Cornell University, JPL, Ball Aerospace, Lockheed Martin,
and the Spitzer Science Center whose tireless dedication have
made success with the IRS and Spitzer a reality.  We dedicate
these first science results to them.  Support for this work
was provided by NASA through an award issued by JPL/Caltech.

\references

\reference{} Armus, L., Heckman, T.M, \& Miley, G.K. 1987, AJ, 94, 831.

\reference{} Blain, A.W., Smail, I., Ivison, R.J., Kneib, J.-P., \& Frayer, 
D.T. 2002, PhR, 369, 111.

\reference{} Chiar, J.E., et al. 2000, ApJ, 537, 749.

\reference{} Decin, L., Morris, P.W., Appleton, P.N., Charmandaris, V., \& Armus, L. 2004, 
ApJ Suppl., in press.

\reference{} Evans, A.S., Mazzarella, J.M., Surace, J.A., \& Sanders, D.B. 2002, ApJ, 580.

\reference{} Genzel, R., Lutz, D., Sturm, E., Egami, E., Kunze, D., 
et al. 1998, ApJ, 498, 589.

\reference{} Goodrich, R.W., Veilleux, S., \& Hill, G.J. 1994, ApJ, 422, 521.

\reference{} Higdon, S.J.U., et al. 2004, PASP, submitted.

\reference{} Houck, J.R., et al. 2004, ApJ Suppl., in press.

\reference{} Huthings, J.B., \& Neff, S.G. 1989, AJ, 97, 1306.

\reference{} Imanishi, M., Dudley, C.C., \& Maloney, P.R. 2001, ApJ, 558, L93.

\reference{} Imanishi, M., Terashima, Y., Anabuki, N., \& Nakagawa, T. 2003, ApJ, 596, L167.

\reference{} Kim, D.C., \& Sanders, D.B. 1998, ApJS, 119, 41.


\reference{} Lonsdale, C.J., Smith, H.E., \& Lonsdale, C.J. 1995, ApJ, 438, 632.

\reference{} Lutz, D., Kunze, D., Spoon, H.W.W., \& Thornley, M.D. 1998, 
A \& A, 333, L75.

\reference{} Lutz, D., Veilleux, S., \& Genzel, R. 1999, ApJ, 517, L13.

\reference{} MacKenty, J.W., \& Stockton, A. 1984, ApJ, 283, 64.

\reference{} Mazzarella, J.M., Gaume, R.A., Soifer, B.T., Graham, J.R., Neugebauer, G.,
\& Matthews, K. 1991, AJ, 102, 1241.

\reference{} Mihos, C.J., \& Hernquist, L. 1996, ApJ, 464, 641.

\reference{} Miller, J.S. \& Goodrich, R.W. 1990, ApJ, 355, 456.

\reference{} Moshir, et al. 1990, IRAS Faint Source Catalog, V2.0.

\reference{} Moutou, C., Verstraete, L., Leger, A., Sellgren, K., \& Schmidt, W. 2000, 
A \& A, 354, L17.

\reference{} Murphy, T.W. Jr., Armus, L., Matthews, K., Soifer, B.T., Mazzarella, J.M., 
Shupe, D.L., Strauss, M.A., \& Neugebauer, G. 1996, AJ, 111, 1025.

\reference{} Reach, W.T., Morris, P., Boulanger, F., \& Okumura, K. 2004, Icarus, in press.

\reference{} Rigopoulou, D., Kunze, D., Lutz, D., Genzel, R., \& Moorwood, 
A.F.M. 2002, A \& A, 389, 374.

\reference{} Rigopoulou, Spoon, H.W.W., Genzel, R., Lutz, D., Moorwood, 
A.F.M., \& Tran, Q.D. 1999, AJ, 118, 2625.

\reference{} Sanders, D.B., et al. 1988a ApJ, 325, 74.
 
\reference{} Sanders, D.B., Soifer, B.T., Elias, J.H., Neugebauer G., \&
Matthews, K. 1988b, ApJ, 328, L35.

\reference{} Shuder, J.M., \& Osterbrock, D.E. 1981, ApJ, 250, 55.

\reference{} Smith, J.D.T., et al. 2004, ApJ Suppl., in press.

\reference{} Soifer, B.T., et al. 1987, ApJ, 320, 238.

Elias, J.H., Lonsdale, C.J., \& Rice, W.L. 1987, ApJ, 320, 238.

\reference{} Soifer, B.T., et al. 2000, AJ, 119, 509.

\reference{} Solomon, P.M., Downes, D., Radford, S.J.E., \& Barrett, J.W. 1997, ApJ, 478, 144.

\reference{} Spoon, H.W.W., Keane, J.V., Tielens, A.G.G.M., Lutz, D., Moorwood, A.F.M., \&
Laurent, O. 2002, A \& A, 385, 1022.


\reference{} Stanford, S.A., Stern, D., van Breugel, W., \& De Breuck, C. 2000, ApJ Suppl., 131, 
185.

\reference{} Strauss, M.A., Huchra, J.P., Davis, M., Yahil, A., Fisher, K.B., 
\& Tonry, J. 1992, ApJS, 83, 29.

\reference{} Sturm, E., et al. 2000, A \& A, 358, 481.


\reference{} Sturm, E., et al. 2002, A \& A, 393, 821.

\reference{} Tran, Q.D., et al. 
2001, ApJ, 552, 527.

\reference{} Veilleux, S., Kim, D.C., Sanders, D.B., Mazzarella, J.M. \& Soifer, B.T.
1995, ApJ Suppl., 98, 171.

\reference{} Veilleux, S., et al. 1997, ApJ, 477, 631.

\reference{} Verma, A., Lutz, D., Sturm, E., Sternberg, A., Genzel, R., \& Vacca, W. 2003, A \& 
A, 403, 829.

\reference{} Voit, G.M. 1992, ApJ, 399, 495.


\end{document}